\documentclass[12pt]{article}
\usepackage[dvips]{color}
\usepackage{mathtext}
\usepackage[koi8-r]{inputenc}
\usepackage{graphicx}
\usepackage[figuresright]{rotating}
\usepackage{amssymb,eucal}
\usepackage{cite}
\date{}

\newcommand{\beq}{\begin{equation}}
\newcommand{\eeq}{\end{equation}}
\newcommand{\beqn}{\begin{eqnarray}}
\newcommand{\eeqn}{\end{eqnarray}}  

\begin{document}
\title{Photon sandwich detectors with WLS fiber readout}

\author{O.~Mineev$^{a,}$\thanks{email: oleg@wocup.inr.troitsk.ru}, E.~Garber$^{b}$, J.~Frank$^{b}$, 
A.~Ivashkin$^{a}$, S.~Kettell$^{b}$, \\ 
M.~Khabibullin$^{a}$, Yu.~Kudenko$^{a}$, K.~Li$^{b}$, L.~Littenberg$^{b}$, \\
V.~Mayatski$^{c}$, N.~Yershov$^{a}$ \\
~\\
$^{a}$Institute for Nuclear Research RAS, 117312 Moscow, Russia \\
$^{b}$Brookhaven National Laboratory, Upton, NY 11973, USA \\
$^{c}$AO Uniplast, 600016 Vladimir, Russia \\}
\maketitle 

\begin{abstract}
A photon detector for BNL experiment E949 is described. 
The detector consists of a lead scintillator ``sandwich'' 
of 25 layers of 5 mm thick scintillator BC404 
and 24 layers of 2 mm lead absorber. Readout is implemented with 
30-60 cm long WLS fibers (BCF~99-29AA) glued into grooves in the BC404. 
Average yield was measured with cosmic rays to be about 43 p.e./MeV. 
Extruded plastic scintillation counters
developed for sandwich detectors of photons for the KOPIO experiment 
are also described.
For a 7 mm thick counter with 4.3 m long WLS fibers spaced at
7 mm a light yield of 18.7 p.e./MeV and time resolution of 0.71 ns
were obtained. A prototype photon veto module consisting of 10
layers of 7 mm thick extruded plastic slabs interleaved with
1 mm lead sheets was tested. 
\end{abstract}
 
\section{Introduction}
BNL experiments KOPIO \cite{e926} and E949 \cite{e949},  to 
search rare kaon decays $K^0_L\rightarrow\pi^0\nu\bar{\nu}$ and 
$K^+\rightarrow\pi^+\nu\bar{\nu}$, respectively,
require extremely high detection efficiency 
of charged particles and photons to suppress the dominant background
from ordinary kaon decays that are 10 orders of magnitude more frequent. 
For this purpose new
photon detectors were designed and built. Their common features
are the wavelength shifter fiber readout and Pb-scintillator sandwich design. 
In this work we report on the E949 collar endcap photon detector and
the KOPIO photon veto prototype module. 

\section{Collar detector for E949}
The E949 collar detectors (CO) are located at the upstream and downstream 
ends of the detector, outside of the endcaps and near the beam axis. 
Their purpose is to detect photons emitted from kaon decays in the kaon
stopping target under small
polar angles. They cover the very forward region
(from $0.995> |cos\Theta| >0.97$) as a part of the E949 photon veto system.

\subsection{Collar design}
\begin{figure}[htpb] 
\vspace{1cm}
\centering\includegraphics[width=13cm,angle=0]{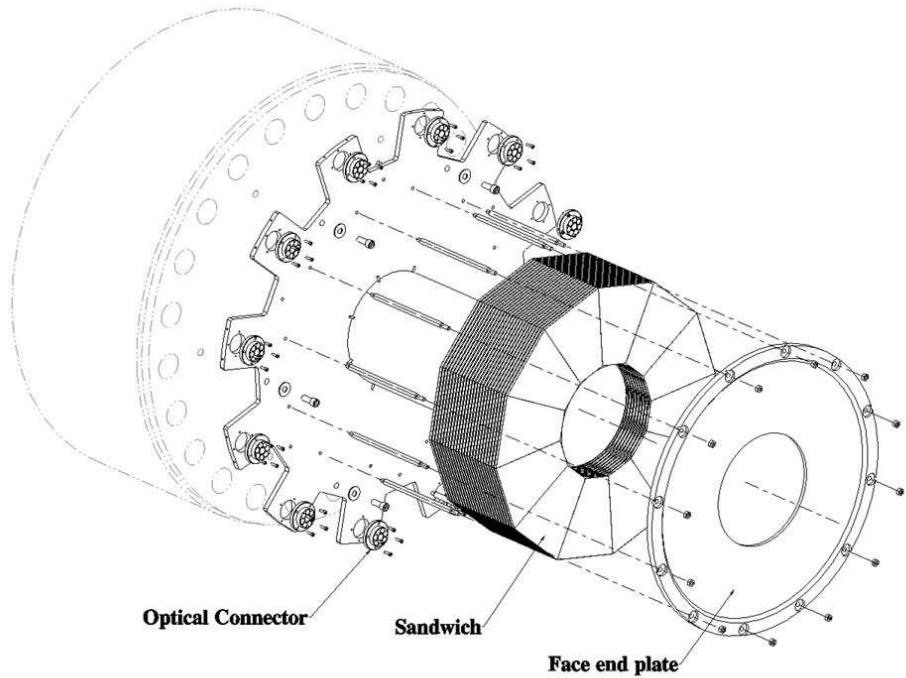}
\caption{Assembly view of the collar sandwich detector.}
\label{fig:collar} 
\end{figure}

The assembly view of downstream CO is shown in Fig.~\ref{fig:collar}.
 The
upstream and downstream CO are similar in design, the difference is in the
geometrical size. The CO consist of 25 layers of 5~mm thick scintillator
BC404 (Bicron, \cite{bicron}) and 24 layers of 2~mm lead absorber interleaved 
by 0.1~mm Tyvek reflector paper. The total radiation thickness is limited by available 
space to $\sim 9X_{0}$.

A single plastic scintillator layer is segmented in 12 wedges (petals)
forming 12 identical azimuthal sectors. The petal length is 165~mm. Light 
from the petals is read out by 16 multiclad wave length shifting (WLS) fibers 
(BCF99-29AA  from Bicron) which are glued into grooves in the scintillator 
to obtain a high light yield~\cite{Ivashkin:sa}. 
The grooves diverge radially along a petal with average spacing of 6~mm. 
The inner end of the fiber is polished and aluminized. 
All 400 fibers for each sector are bent under angle of 90$^{\circ}$ 
and terminated in a connector. 
Fiber length is variable with an average of about 45~cm.

Each absorber layer is composed of
two halves of a 2~mm thick lead semicircular annular disk. The Pb layers are laid out
with a rotation angle of 30$^{\circ}$ relative to the neighboring layer,
providing coverage of the cracks between the Pb halves in subsequent layers.
The gaps between scintillator petals are also staggered so that there is
no projective cracks for photons over the CO face.

The sandwich assembly is compressed between a base plate and an endplate
disk by 12 standoffs. The standoffs prevent the collar elements from
shifting down due to gravity.

\subsection{Construction and assembly}
The scintillator petals were cut from cast sheets of BC404 scintillator with 0.1~mm tolerance.
Since the sheet has significant thickness variation (range from 4.8 to 5.2~mm),
petals were hand selected in groups of similar thickness. The petals
from the same group were laid out to form a single layer. 

Sixteen symmetrically spaced U-shaped grooves were cut along the petal 
and polished. Gluing was done in two steps. First, the grooves were filled
with BC-600 epoxy, and the 65~cm long fibers were inserted. 
In the second step,
the glue was applied over the fibers to fill any remaining gaps.

The scintillator is etched in a process that results in the formation of a 
micropore reflector deposit over the surface.
Long term effect of Pb diffusion in the chemical reflector is unknown
so Tyvek sheets were used to separate the scintillator and lead. The
Tyvek also works as a reflector over the fibers.

\begin{figure}[htpb]
\centering\includegraphics[width=10cm,angle=0]{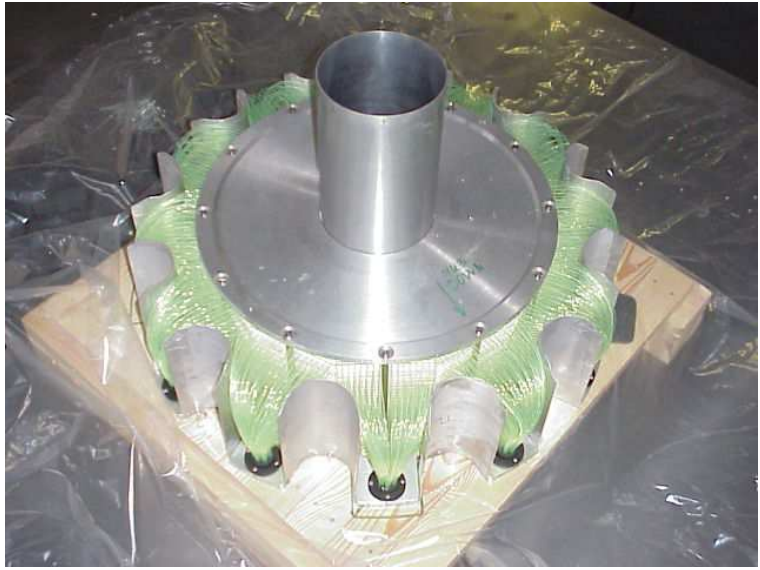}
\vspace{1cm}
\caption{The collar detector before installation in the set--up. After
installation the fibers have been shielded by protective cover.}
\label{fig:collar_photo} 
\end{figure}

The most critical point in the collar assembly is the insertion of the fibers
in the connector. Before gluing all fibers were visually inspected and then 
bent over a cylinder with diameter of 5~cm to expose cracks. The failure 
rate (a fiber is broken) is below 0.5\%.
If Bicron fiber was bent with a radius of less than 3~cm,
the crazing of the cladding was observed to develop in the curved part for
some fibers. This process takes place in a hour and continues over 24--48 hrs.
After that stabilization is observed (no additional crazing after 48 hours).
The fibers were routed to a connector with a bending radius of 4~cm or larger.
Even so some fibers show crazing which is not harmful as long as the
number of such fibers is less than 1\%. 
Then the fibers have been glued in the connector, the ends were cut and polished.
A view of the assembled collar detector is shown in Fig.~\ref{fig:collar_photo}.
\subsection{Cosmic test}
The detector was tested with cosmic rays.
The trigger counters were placed above and below each sector. Though the area
of the trigger counters is larger than that of a sector, the 50~cm separation
selects minimum ionizing particles (MIP) with incident angles close to normal.
A MIP deposits about 25~MeV of visible energy in the detector.
A phototube is attached to the fiber connector through a silicone cookie.
All sectors were measured with the same photomultiplier EMI-9954 which has
a photocathode sensitivity extended in the green light region.

\begin{figure}[htpb]
\vspace{1cm}
 \centering\includegraphics[width=14cm,angle=0]{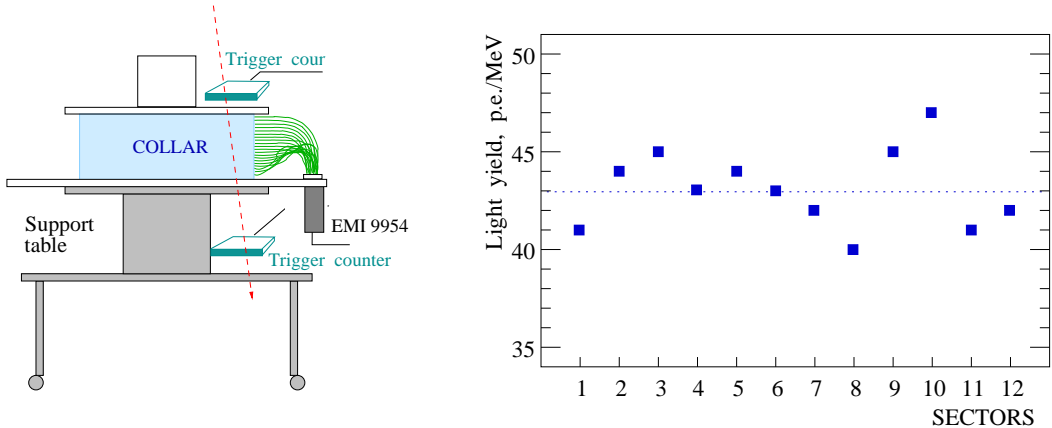}
 \vspace{0.5cm}
\caption{Light yield of the collar detector for each sector. Test bench
is also shown.}
\vspace{0.5cm}
\label{fig:collar_ly} 
\end{figure}

The average number of detected photoelectrons (p.e.) is calculated by dividing
the MIP peak position by the single electron peak position. An attenuator
was used to scale the large MIP signal to the single electron amplitude.
The cosmic test results are presented in Fig.~\ref{fig:collar_ly}.
Average light yield is about 43~p.e./MeV, where MeV is visible energy.
Deviation from this value in different sectors is $\pm$3~p.e., this is
within the accuracy of the single electron peak calibration.

\section{Sandwich veto detector for KOPIO experiment}
\subsection{General description}
One of the most important  parts of the BNL experiment KOPIO is the determination that 
nothing other than single $\pi^{0}$ was emitted in the decay 
$K^0_L\rightarrow\pi^0\nu\bar{\nu}$, i.e. a high efficiency veto of any extra
particles. It requires an extremely high photon detection efficiency of 
better than 0.9996 per photon, without excessive loss from random vetoes. 
High light yield over all detector volume and good timing are needed to 
satisfy these requirements.

\begin{figure}[htpb]
\centering\includegraphics[width=13cm,angle=0]{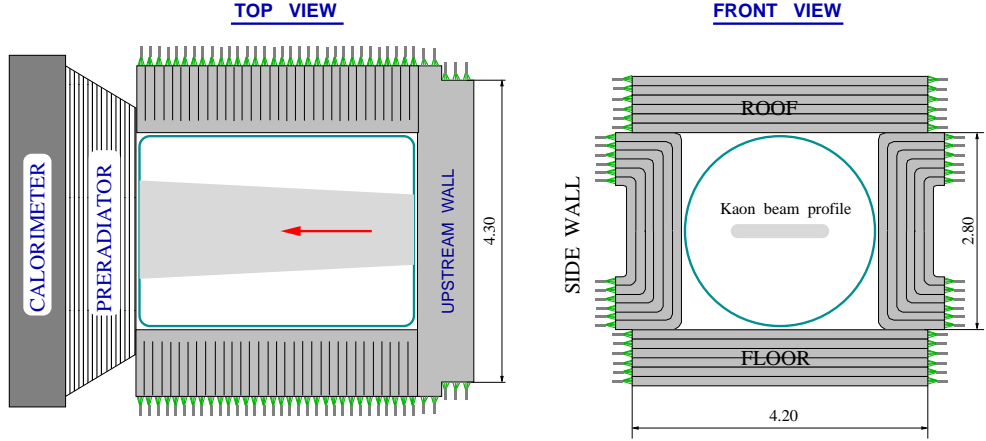}
\vspace{1cm}
\caption{The schematic view of the barrel veto detector in the KOPIO.}
\label{fig:barrel_view}
\end{figure}

Lead--plastic scintillator sandwich detector is being considered for the
barrel photon veto in KOPIO. To read light from the scintillators,
an embedded WLS fiber readout will be used. Similar detectors were designed
for BNL experiment E787~\cite{Atiya:1992vh} (without WLS fibers) and 
KEK experiment E391~\cite{e391}.  Thickness of the lead sheets
is considered to be 0.5--1.0~mm, and the thickness of extruded polystyrene
scintillator slabs is 7~mm. WLS fibers are glued in the grooves which run
along the scintillator slabs with spacing of 10~mm. The schematic view of the
barrel veto is shown in Fig.~\ref{fig:barrel_view}. 

\subsection{Sandwich module design}
The results of tests of single extruded counters and a sandwich module
prototype are reported in Ref.~\cite{Kudenko:qj}. The main results for extruded 
counters with WLS fiber readout are summarized in Table~\ref{table:counters}.
\begin{table}[htbp]
\caption{ Parameters of  extruded polystyrene counters with 4.3~m long WLS fiber readout.
Fibers: multi--clad BCF99-29AA and single--clad BCF92 of 1~mm diameter.
 A counter made of BC404 scintillator is also shown for comparison.}
 \begin{center}
 \hspace{1cm}
 \begin{tabular}{ccccc}  
 \hline
 &&&&\\
 Counter thickness  & Spacing  & Fiber type   & Light yield   & $\sigma_{t}$ \\
       mm           &    mm    &              &  p.e./MIP     &     ns          \\
 &&&&\\
 \hline
 &&&&\\
7  &   19                      & multi--clad &    11.2       &                 \\
7  &   10                      & multi--clad &    19.6       &       0.85      \\
7  &   10                      & single--clad&    14.4       &       0.87      \\
7  &   7                       & multi--clad &    26.2       &       0.71      \\
7  &   7                       & single--clad&    20.8       &       0.76      \\
3  &   10                      & multi--clad &     8.5       &       0.92      \\
&&&&\\
\hline
&&&&\\
7 (BC404)& 7              & multi--clad &    32         &       0.65      \\
&&&&\\
\hline
 \end{tabular}
 \end{center} 
 \label{table:counters}  
\end{table} 
 
Extruded polystyrene scintillator with WLS fiber readout produces 0.8 of
light yield of BC404 scintillator. Single--clad (s.c.) and multi--clad (m.c.) fibers
provide practically the same time resolution though in the tested configuration 
the light yield for m.c. fibers was 20--24\% higher than that for s.c. ones.

\begin{figure}[htpb]
\centering\includegraphics[width=13cm,angle=0]{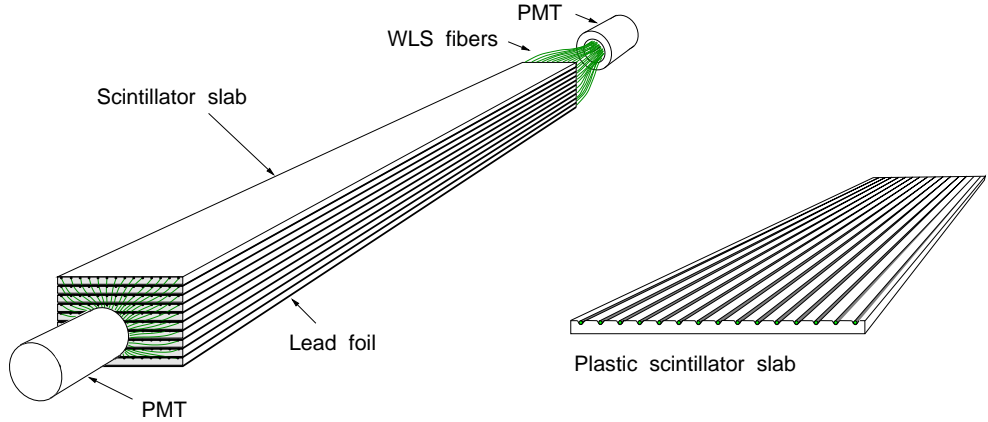}
\vspace{1cm}
\caption{The schematic view of a sandwich module.}
\label{fig:module_view} 
\end{figure}

\begin{figure}[htpb]
\centering\includegraphics[width=13cm,angle=0]{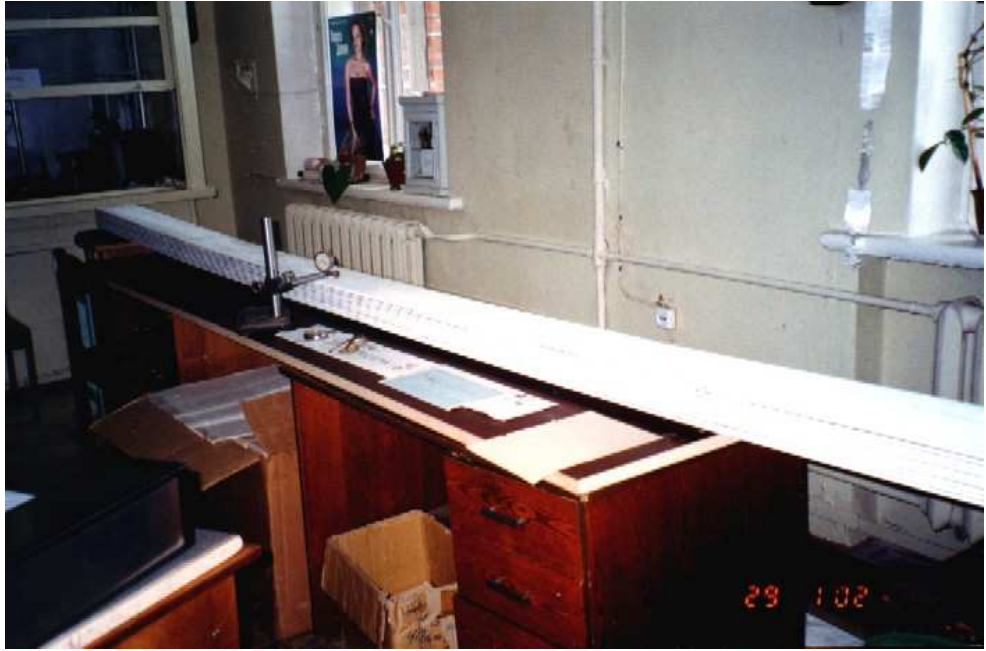}
\vspace{1cm}
\caption{Photo of the 4~m long glued module under mechanical test.}
\label{fig:sag} 
\end{figure}

A schematic view of a sandwich element (module) which is used to build roof and
floor parts of the barrel is shown in Fig.~\ref{fig:module_view}. 
The number
of lead--plastic layers in a single module is 15. The width is 130~mm, spacing
between grooves is 10~mm, groove depth is 1.5--2.0~mm. Sandwich module is 4.2~m
long without extending fibers. A phototube views a bundle of 195 WLS fibers at
each end providing a both--side readout. A chemical etched 
reflector~\cite{Kudenko:qj}
is used to cover the scintillator slabs. The scintillator and lead are glued together
in a monolithic block using an elastic polyurethane glue with high viscosity. The
glue does not soak in the micropore chemical reflector. After gluing the lead plates
with scintillator covered by the reflector we found the light output reduction
of about 6\%, but then no degradation in the light yield was observed for two months.
To test the mechanical properties, a 4~m long dumb module (no fibers) had been
assembled (see Fig.~\ref{fig:sag}).
10 layers of scintillator and lead were coupled 
together only by glue.
An additional layer of 0.1~mm stainless steel foil was also glued at the first
layer of the plastic. The module sag under own weight was measured to be 4~cm
after two weeks of test. This value is close to the expected sag if the module
would be a solid polystyrene.

\begin{figure}[htpb]
\centering\includegraphics[width=13cm,angle=0]{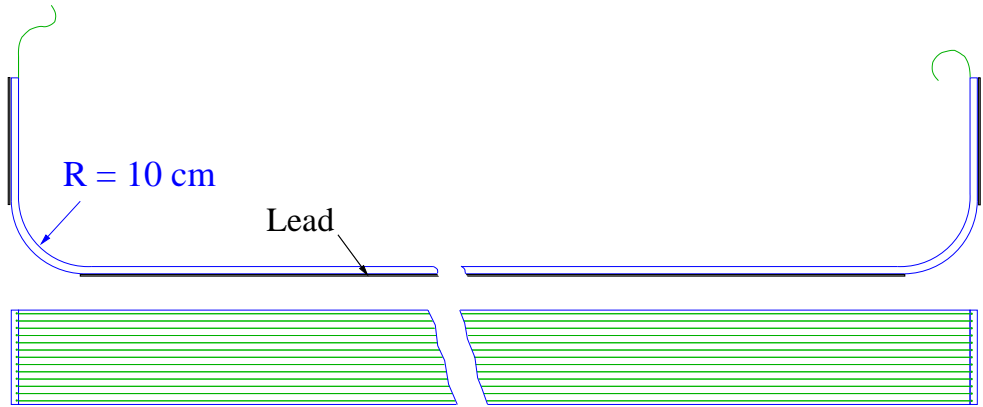}
\vspace{1cm}
 \caption{The view of a single layer for the side wall.}
\label{fig:bend_slab} 
\end{figure}

Unique feature of the veto side wall is the bent sandwich modules assembled of
bent scintillator slabs as shown in Fig.~\ref{fig:bend_slab}.
Bending with 10~cm radius can be done heating the scintillator slabs with fibers
while the optical glue has not hardened yet. This technique was successfully
tested.

\begin{figure}[htpb] 
 \centering\includegraphics[width=12cm,angle=0]{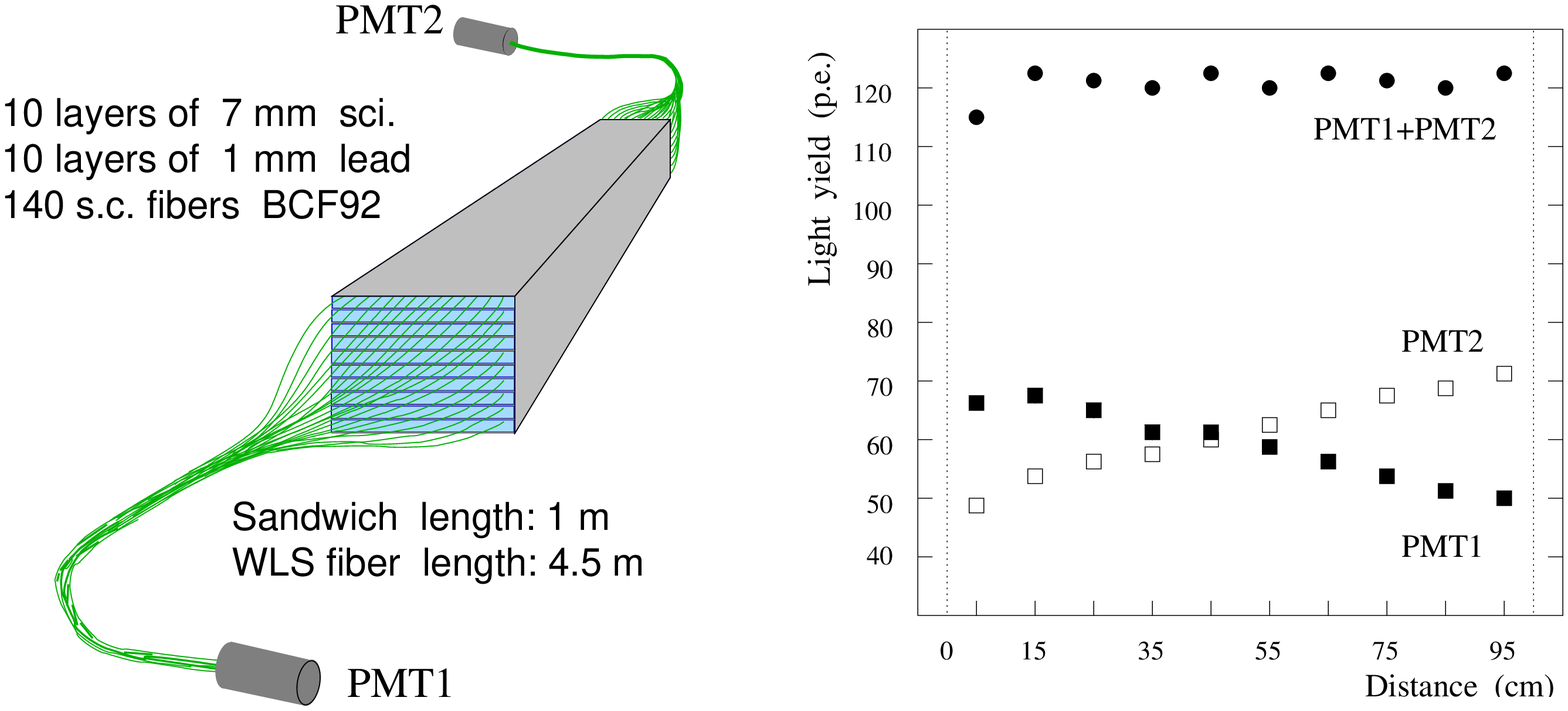}
 \vspace{1cm}
\caption{Light yield of the sandwich module prototype at cosmic ray test.}
\label{fig:sandwich_proto} \end{figure}

The sandwich prototype module was made with a length of 1~m while the length
of the WLS fibers was 4.5~m, i.e. close to length in the real detector.
Single--clad BCF-92 fibers were glued with 10~mm spacing.
Light yield was measured to be 122~p.e. for MIP, and time resolution was 360~ps (rms).
Light yield nonuniformity along the module was measured in 10~cm steps.
As shown in Fig.~\ref{fig:sandwich_proto}, 
the summed light yield from both ends
was stable with deviation $\pm1$\% from the average over 90~cm length.

\subsection{Evaluation of sandwich veto performance}
The expected performance of the  barrel veto is summarized in 
Table~\ref{table:performance}.
\begin{table}[h]
\caption{Expected  barrel veto performance.}
 \begin{center}
\hspace{0.2cm} 
 \begin{tabular}{|l|c|l|}  
 \hline
 &&\\
 ~~~~~Parameter  &       Value             &       ~~~~~~~Note  \\ 
 &&\\                
 \hline 
 &&\\
Integral inefficiency        &   $<4\times10^{-4}$     &    Based on tests and E787 results    \\
$\sigma_{E}$/$\sqrt{E(GeV)}$, ~~\%&            3.6          &    Sampling fluctuations from GEANT   \\
$\sigma_{t}$/$\sqrt{E(GeV)}$, ~~ps&             63          &    Stochastic term from prototype test \\ 
Position resolution, cm    &              3.8          &    Veto segmentation and timing\\
Angular resolution          &                          &    \\
for 100~MeV photon, mrad    &             250          &  GEANT simulation     \\    
&&\\
\hline
 \end{tabular}
 \end{center} 
 \label{table:performance}  
\end{table}  
Visible energy fraction for 1~mm lead and 7~mm scintillator is about 0.39.
From test results we evaluate the photoyield to be 4.0~p.e. per MeV of
the energy deposited in both active (scintillator) and passive (lead) layers.
It means we can collect 400 p.e. for a typical 100~MeV photon.
Extrapolating the parameters obtained for the prototype module, the time resolution 
of the veto detector is estimated to be close to 200~ps (rms) for a 100~MeV 
photon.  Direct extrapolation for higher energies  gives 63~ps/$\sqrt{E_{\gamma}[GeV]}$.
However, at 1~GeV the time resolution includes not only stochastic term but also the
systematic constant term, which is difficult to estimate.  

Since the light propagation velocity in the 
fibers was measured to be 17~cm/ns, the accuracy of a localization  of an electromagnetic 
shower could be about 3.4~cm (rms) along the fibers for a 100~MeV photon. 
The resolution in the other direction is defined by the width of the modules. 
For a 130~mm width we have $\sigma_{x}$=3.8~cm. 

We are able to recover the shower direction measuring the center-of-gravity 
at each level and drawing the line through these centers by $\chi^{2}$ method. 
This method  gives the systematic shift which was corrected.  We simulated
the photons irradiated from the center of the decay region under angles in uniform 
populated range from 45 to 135$^{\circ}$.  Angular resolution of 244~mrad 
($E_{\gamma}$=100~MeV) for a typical angle of 70$^{\circ}$ leads to the measurement
of the $K^0_L$ decay position with an accuracy of about 40~cm.

\section{Conclusion}
Photon veto detectors for BNL experiment E949 were manufactured and 
tested with cosmic rays. 
The detectors consist of 25 layers of 5 mm thick scintillator BC404 
and 24 layers of 2 mm lead absorber. Readout is implemented with 
multi--clad WLS fibers (BCF~99-29AA) of 30-60 cm length. The average yield was measured with
cosmic rays to be about 43 p.e./MeV.

A lead--plastic scintillator sandwich veto system is designed for  BNL experiment
KOPIO.  The barrel veto features high light output and
readout segmentation over its thickness. To read light from the scintillators, 
an embedded  wavelength--shifting fiber readout technique will be used. 
A prototype photon veto module consisting of 10 layers of 7~mm thick extruded 
plastic slabs interleaved with 1~mm lead sheets was tested. Single--clad 
BCF--92 fibers of 4.5~m length spaced at 10~mm were used for the readout.
The module yielded 122~p.e. per MIP and time resolution of 360~ps.




\begin{thebibliography}{99}
\bibitem{e926}I.-H.~Chiang et al., BNL Proposal E926, Measurement of 
$K_L\rightarrow\pi^0\nu\bar{\nu}$, September 1996.
\bibitem{e949} Brookhaven AGS Experiment 949 (http://www.phy.bnl.gov/e949/)
\bibitem{bicron} Bicron Corporation, 12345 Kinsman Rd., Newbury, OH 44065 USA.
\bibitem{Ivashkin:sa}
A.P.~Ivashkin {\it et al.}, 
Nucl.\ Instrum.\ Meth.\ A {\bf 394}, 321 (1997).
\bibitem{Atiya:1992vh}
M.S.~Atiya {\it et al.},
Nucl.\ Instrum.\ Meth.\ A {\bf 321}, 129 (1992).
\bibitem{e391} K.~Abe {\it et al.}, KEK-Preprint-2000-89, August 2000.
\bibitem{Kudenko:qj}
Y.G.~Kudenko {\it et al.},
Nucl.\ Instrum.\ Meth.\ A {\bf 469}, 340 (2001).
\end{thebibliography}
\end{document}